# Supramolecular Thermo Aero-able Gelators (STAGs) for synthesis of hydrogels

Feng Feng Xue,[a,b] Dan Dan Yuan,[a] Atharva Sahasrabudhe,[a,b] Subharanjan Biswas,[b] Peng Wang,[a] Xiao-Yan Tang,[a] Dianyu Chen,[a] Rongxin Yuan,*[a] and Soumyajit Roy*[a,b]

Supramolecular Thermo Aero-able Gelators (STAGs): Tartaric acid, urea, and guanidine with carboxylic amide and imine moieties respectively as supramolecular synthons are introduced to cross-link and aerate ('aero-able') polyacrylate networks for synthesis of hydrogels. They are bi-functional hence present a greener alternative to the existing cross-linkers and gelators.

## Introduction

Syntheses of super-swelling hydrogels continue to be a fascinating field.[1,2] Courtesy: a diverse array of applications, ranging from water retention hygiene products, space-suits, surgical wipes to artificial vegetation, controlled release of drugs and pesticides and beyond. Synthesis of such hydrogels involves two key steps: cross-linking of the polymer and blowing or aeration of the cross-linked polymeric network to facilitate water-retention by the polymeric network. In the wake of green chemical requirements, design of a reagent that is bi-functional and can hence both cross-link and aerate the network would be a welcome step forward. Supramolecular Thermo Aero-able Gelators (from here onward abbreviated as STAGs) like tartaric acid (used as L-tartaric acid), urea and guanidine (used as guanidinium chloride), reported here is a step forward in that direction. [Here onwards they are abbreviated as TA, U and GD respectively.] They are bi-functional. They incorporate two functional facets – of cross-linking and aerating – in the guise of one molecule. This work aims to exploit aspects of supramolecular cross-linking and thermal degradation mediated aeration under one tenet of the molecules that are well-documented in the literature and thus prompted the choice of STAGs. In fact, GD is used here to compare and contrast the STAG activity of the other two molecules, TA and U. GD has more stable thermal profile than the other two but shares the supramolecular hydrogen bonding capacity like that of TA and U and hence can act as a comparative standard for the other STAGs. The concept of supramolecular cross-linking is well-known in the literature. For instance, bis-thymine units have been used to supramolecularly cross-link diamino-pyridine functionalized co-polymers forming thermo-responsive gels.[3] Stimuli-responsive metallo-supramolecular polymers,[4] switchable hydrogels,[5] supramolecular cross-linked $C_{60}$ tagged with polymers for enhanced mechanical properties have also been reported recently.[6,7] Supramolecular cross-linking for encapsulation of proteins and their release have also been demonstrated.[8] Co-ordinatively cross-linked supramolecular polymers, organo-gels prepared by 'click chemistry' in a supramolecular environment,[9] supramolecular polymers[10] and even constitutionally dynamic polymers ("dynamers")[11,12] have been reported and extensively studied. The capacity of supramolecular cross-linking in directing structures of extended oxometalate based macromolecules using U and GD has also been shown.[13] Here we take the next step. We exploit the capacity of TA, U and GD to supramolecularly cross-link polymeric matrix of polyacrylate and additionally to create space for incumbent water molecules by thermally aerating a fraction of the network. (Fig. 1)

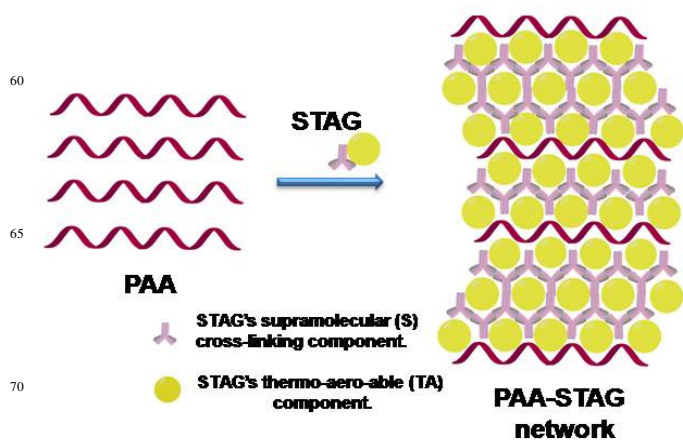

**Fig. 1** Schematic representation of STAGs: STAGs are schematically shown to serve both supramolecular (S) cross-linking as well as thermal aeration (TA) function in PAA-STAG network.

## Experimental

### Synthesis of the starting low MW polyacrylic acid

Acrylic acid used in this study was reagent grade. All other chemicals were analytical grade. The reagents were purchased from J & K Scientific Ltd. China. Tartaric acid was used in enatiomarically pure form as L-tartaric acid. The reagents were used as obtained without further purification. 60g deionized water and 2g sodium hydrogen sulfite were added into a reactor, and the mixture was stirred until all the sodium hydrogen sulfite was dissolved. The reactor with the contents then was slowly heated in an oil-bath and the temperature was raised to 60 °C. Later a mixture of 20g acrylic acid, 0.2g ammonium peroxydisulfate and 20mL deionized water was mixed and was fed into the reactor containing 60g deionized water and 2g sodium hydrogen sulfite by dropping the whole contents over 40 minutes. The reactor with the contents was stirred at 60 °C for another 1.5 hours. Later the reactor was cooled and the polyacrylic acid so obtained was used further for the swelling and the cross-linking experiments as described in the text. Here onward polyacrylic acid is referred to as PAA.

**Swelling experiments**

The prepared PAA polymer in this experiment is used as a starting material for synthesis of supramolecularly cross-linked hydrogels. 4 g of prepared polymer (PAA) was mixed with a range of different concentrations (5% - 55% w/w) of TA, U and GD. The STAGs (TA, U, and GD) were added as solids to the PAA solution prepared. The upper limit of this mixing was dictated by the physical ability of the polymeric solution in water to dissolve the STAGs. After mixing, the mixture was heated to 150 °C for 30 minutes in an oven with continuous air-flow. The resulting dry-material was taken out and was subjected to various analyses, including the equilibrium swelling experiments.[14] The equilibrium swelling experiments were carried out by using each time 0.5 g (say, $W_1$ g) of the heated, dried, crushed and sieved polymeric materials (grain size ca. 100μm) by allowing it to swell at 37 °C in a buffer solution of pH 6 for 5 days. The use of buffer solution is to avoid the effects of ionization of PAA on swelling. After swelling was complete, the sample was dried in air and weighed again ($W_2$ g). The ratio of $W_2/W_1$ was taken as the equilibrium swelling ratio, represented from here onwards as Q.

**Spectroscopy and microscopy experiments**

FTIR spectra reported in this study were recorded as KBr pellets with a Nicolet 380 FT-IR spectrometer in the range of 4000–500 cm$^{-1}$. SEM images were acquired by Hitachi Tabletop Microscope, Model No. TM3000. $^1$H NMR spectra were recorded in 400 MHz Brucker Avance Spectrometer. UV-Vis spectra were recorded in the range 200 - 1000 nm with a Shimadzu UV-160A spectrophotometer and evaluated with a program associated with the spectrometer.

**Thermogravimmetric experiments**

Thermogravimmetric analysis was performed using TGA/SDTA851e, Mettler Toledo, Switzerland. For the TGA experiments ca. 10 mg of dried materials were heated from 298 to 778 K with a rate of 15.00 K/min, under an oxygen flow of 10.00 mL/min. The weight loss (ΔW) reported here was determined under these conditions at 150 °C. Correspondingly swelling values at different weight losses were calculated and have been analyzed here to find the influence of thermal decomposition of STAGs in blowing of the polymeric networks.

**Size exclusion chromatography experiments**

Size exclusion chromatography was carried out using tetrahydrofuran (THF) as the mobile phase with Waters 717 Autosampler and Waters 2410 refractive index detector. The number average molecular weight ($M_n$) and polydispersity index (PDI) of the synthesized PAA was obtained using this method. The PDI of the synthesized PAA was found to be less than 1.2.

**Preparation of xerogels for X-ray diffraction experiments**

The xerogels were obtained by freeze-drying the corresponding hydrogels in an Eyela FDU 1000 freeze drier at a temperature of -60 °C for 5 hours.

**X-ray diffraction experiments**

The XRD patterns were obtained by Rigaku (mini flex II, Japan), using CuK$_α$ radiation having an incident wavelength of 1.541 Å operating under a voltage of 30 kV and a current of 50 mA. The scan rate was 12 °/minute.

**Optical diattenuance and retardance experiments**

In order to augment the structural investigation of the hydrogels we further undertook diattenuation and retardance study. Here we describe very briefly how the polarization properties of light scattered from samples may give additional micro-structural functional information on complex molecular anisotropy/organization. This information is otherwise hidden in polarization blind light scattering-based measurements. In conventional polarimetry, the polarization information is quantified in terms of the conventionally defined degree of polarization of light. Performing simple measurements of co-polarized and crossed-polarized light intensity using a given input polarization state of incident light as standard or by recording selected Stokes vector elements such measurements are done (4 × 1 intensity vector describing the polarization state of light, defined afterwards).[15] Such semi-quantitative information so obtained on the polarization properties of the sample are then extracted by employing empirical formulations on the selected polarimetry measurements. Simpler traditional polarimetry analyses are though suitable for optically clear media, complexities arise when several polarization events are exhibited in the sample simultaneously. It is even more complicated when the probe in question itself scatters light as is the case for the given PAA and PAA-STAG hydrogels investigated in this study.[16] Simultaneous superposition of several polarization effects lead to a complex interrelation that therefore represent several *lumped* effects with much *inter-element cross talk*. This masks potentially interesting polarization metrics and hinders their unique interpretation. Hence for extraction, quantification and unique interpretation of individual, intrinsic polarimetry effects of such complex systems, like the hydrogels in question, a more quantifiable and generally applicable technique is thus required. One such method is Mueller matrix decomposition approach, which aims to solve this problem. This method extracts constituent polarization properties from a given 'lumped' system's Mueller matrix of any unknown complex system (summarized below). We use this methodology to analyze PAA and PAA-STAGs hydrogels in this question. As model systems PAA, PAA-STAG-TA and PAA-STAG-U were used and their structures in microscopic domains determined using this method. Since this method is novel we describe Mueller matrix formulation in the context of Stokes vectors a bit more in details, since understanding is needed to analyze these structural investigations.

The interaction of polarized light with any medium can be well described either via Jones or Stokes – Mueller matrix algebra.[15,16] The latter is preferred while dealing with polarization losses (depolarizing interactions). Such a situation has been encountered in our samples as they are all scattering media. In the Stokes – Mueller formalism, the polarization state of light is represented by four measurable quantities (intensities). These are known as the Stokes parameters.[15,16] Such four Stokes parameters of light (I, Q, U and V, elements of the Stokes vector **S**) are represented by six intensity measurements which are performed with ideal polarisers. 'I' is the total detected light intensity and corresponds to the addition of the two orthogonal component intensities. 'Q' is the difference in intensity between horizontal (0°) and vertical (90°) polarization states. 'U' presents the portion of the intensity that corresponds to the difference in intensities of linear + 45° and 135° polarization states, whereas, 'V' is the difference between intensities of right circular and left circular polarization states.

The Stokes vectors represent the polarization state of light. It is a 4 × 4 matrix **M,** known as the Mueller matrix and it describes the polarizing transfer function of any medium as it interacts with light (as given in equations 1 & 2).[15,16]

$$S_o = M\ S_i \quad (1)$$

$$\begin{bmatrix} I_0 \\ Q_0 \\ U_0 \\ V_0 \end{bmatrix} = \begin{bmatrix} m_{11} & m_{12} & m_{13} & m_{14} \\ m_{21} & m_{22} & m_{23} & m_{24} \\ m_{31} & m_{32} & m_{33} & m_{34} \\ m_{41} & m_{42} & m_{43} & m_{44} \end{bmatrix} \begin{bmatrix} I_i \\ Q_i \\ U_i \\ V_i \end{bmatrix} = \begin{bmatrix} m_{11}I_i & m_{12}Q_i & m_{13}U_i & m_{14}V_i \\ m_{21}I_i & m_{22}Q_i & m_{23}U_i & m_{24}V_i \\ m_{31}I_i & m_{32}Q_i & m_{33}U_i & m_{34}V_i \\ m_{41}I_i & m_{42}Q_i & m_{43}U_i & m_{44}V_i \end{bmatrix} \quad (2)$$

where $S_i$ and $S_o$ stand for the output and the input Stokes vectors of the light, respectively. The medium's polarization properties are encoded in the above sixteen elements of the 4 × 4 real Mueller matrix **M.** This matrix can thus be thought of as the complete 'optical polarization fingerprint' of a sample.[15,16] In the following section we describe how inverse polarimetry analysis is performed using decomposition of Mueller matrix.

Decomposition of Mueller matrix aims to solve the inverse problem. In other words it extracts constituent polarization properties from a given 'lumped' system Mueller matrix of any other unknown complex system. The method consists of decomposing the given Mueller matrix **M** into a product of three 'basis' matrices, (equation 3),[16-18]

$$M \Leftarrow M_\Delta \cdot M_R \cdot M_D \quad (3)$$

where '⇐' symbol is used to signify the decomposition process. Here, the matrix $M_\Delta$ describes the depolarizing effects of the medium. $M_R$ takes care of the effects of linear and circular retardance (or optical rotation). On the other hand, $M_D$ includes the effects of linear and circular diattenuation. Now using the known forms of these 'basis' matrices, the measured **M** can be decomposed. Such decomposition is done through a series of manipulations into the three 'basis' matrices. This product decomposition process was developed by Lu and Chipman for optically clear media.[17]

Recently this method has been extended to encompass several complex scattering media exhibiting simultaneous multiple polarization and scattering effects.[16,18] The ability of this approach to delineate individual intrinsic polarimetry characteristics is also validated on such composite systems. Moreover complex turbid media such as biological tissues have also been structurally investigated using this method. Once calculated, the constituent basis matrices are further analyzed to obtain individual polarization medium properties. Specifically, $d$, $\Delta$, $\delta$, and circular retardance or optical rotation ($\psi$, circular retardance = 2×optical rotation), are determined from the decomposed basis matrices. The magnitude of diattenuation ($d$) is determined from $M_D$ using the following relation,[16-18]

$$d = \frac{1}{M_D(1,1)} \sqrt{M_D(1,2)^2 + M_D(1,3)^2 + M_D(1,2)^2 + M_D(1,4)^2} \quad (4)$$

Here, the coefficients $M_D(1,2)$ and $M_D(1,3)$ stand for linear diattenuation for horizontal (vertical) and + 45° (- 45°) linear polarization respectively. The coefficient $M_D(1,4)$ represents circular diattenuation. As a next step, depolarization is quantified through the depolarization matrix $M_\Delta$ as the net depolarization coefficient Δ (as mentioned in equation 5),

$$\Delta = 1 - \frac{|tr(M_\Delta) - 1|}{3}, 0 \leq \Delta \leq 1 \quad (5)$$

The magnitudes of linear retardance $\delta$ and optical rotation $\psi$ is calculated from the various elements of the decomposed retarder matrix $M_R$ using following relations (as given in equations 6 & 7),

$$\delta = cos^{-1}\left(\sqrt{\left(M_R(2,2) + M_R(3,3)\right)^2 + \left(M_R(3,2) - M_R(2,3)\right)^2} - 1\right) \quad (6)$$

and

$$\psi = tan^{-1}\left(\frac{M_R(3,2) - M_R(2,3)}{M_R(2,2) + M_R(3,3)}\right) \quad (7)$$

The details of our experimental spectral Mueller matrix polarimetry system used for this study have been reported previously elsewhere.[19] Here we summarize the experimental set-up that was used for these measurements. The system comprises of a collimated output from a Xe lamp (HPX-2000, Ocean Optics, USA) as excitation source, a polarization state generator (PSG) and a polarization state analyzer (PSA) unit to generate and analyze polarization states required for 4 × 4 sample Mueller matrix measurements. This is coupled to a spectrometer (USB

4000FL, Ocean Optics, USA) for spectrally resolved (wavelength $\lambda$ = 500 - 750 nm, resolution ~ 1.8 nm) signal detection. The PSG unit comprises of a fixed linear polarizer ($P_1$, polarization axis oriented at horizontal position) followed by a rotatable broadband quarter waveplate ($Q_1$) which is mounted on a computer controlled rotational mount. The sample-scattered light collected (at user selected detection angle $\gamma$) and collimated using an assembly of lenses, is then passed through the PSA unit. This is finally recorded using the spectrometer. The PSA unit essentially consists of a similar arrangement [a fixed polarizer $P_2$ at vertical position (cross-state) with respect to polarizer $P_1$ and a rotatable broadband quarter waveplate $Q_2$] as that of the PSG, but it is positioned in reverse order. A sequence of sixteen automated measurements are performed by changing the orientation of the fast axis of the quarter waveplates ($\theta$) of the PSG unit (for generating the four required elliptical polarization states) and also that of the PSA unit (for analyzing the corresponding polarization states). The orientation angles ($\theta$ = 35°, 70°, 105° and 140°) are chosen based on optimization of the PSG and PSA matrices to yield stable system Mueller matrices. These sixteen spectrally resolved intensity measurements are then combined to yield the sample Mueller matrices for the wavelength range 500–750 nm. The performance of this Mueller matrix polarimeter is calibrated using Eigen value calibration method. It also yields the actual values of the system PSG and PSA matrices at each wavelength. It hence allows the correction of non-ideal polarization elements and their wavelength response. The system is completely automated and is capable of Mueller matrix measurement with high accuracy (elemental error < 0.01). It spans the entire spectral range 500 – 750 nm. Mueller matrices in the wavelength range 500 – 750 nm are recorded from the hydrogel samples in the transmission geometry (detection angle $\gamma$ = 0°). The recorded Mueller matrices are then subjected to the polar decomposition analysis to yield quantitative polarimetry characteristics, viz., $d$, $\delta$, $\psi$ and $\Delta$ of the samples.

## Results and Discussions

### Analyzing effect of STAGs on swelling, molecular weights and cross-linking density

Both these experiments clearly indicate the influence of TA, U and GD on swelling patterns and MWs of the polymeric materials synthesized. The highest Q for pure PAA was 1.25 whereas that of with TA was 16.4, with U, 3.0 and with GD was ca.1. The above observations indicate the efficacy of STAGs in cross-linking. It is further observed that Q values also altered with a change in the concentration of the STAGs. (Fig 2a) In case of TA it is observed that Q increases to a maximum at 5% (w/w) TA (or 0.033 mole-fraction of TA, from here onwards written as $x_{TA}$) whereas in case of U the maximum is reached at 10% (w/w) U (or 0.166 by mole-fraction of U, from here onwards written as $x_U$). $x_{TA}$ and $x_U$ represent the optimum mole-fraction of TA and U STAG respectively for maximum swelling. The overall trend of variation of swelling ratio (Q) with concentration of STAGs in both the cases of TA and U shows a 'ʌ' type trend, i.e., from low mole-fraction of TA and U, Q increases and reaches a maximum till $x_{TA}$ and $x_U$ and then tails off to a low Q as TA and U mole-fraction is further increased. GD in general shows a decrease in the Q value with an increase in concentration. Likewise, the MWs of the polymeric materials cross-linked by TA, U and GD showed correspondingly similar patterns. For instance, the MW of the pure PAA synthesized is 5.3 kDa, whereas those of PAA with TA (at $x_{TA}$), U (at $x_U$), at points of highest Qs or maximum swelling are 18.6 and 13.9 kDa respectively. With GD on the other hand MW, at almost all mole-fractions, is constant ca. 14.3 kDa, consistent with our observation of an almost constant Q value at various mole-fractions. Now it is perhaps apt to analyze the results a bit more closely. To do so, we first wanted to find the $M_c$ (the MW of the polymeric units between two successive cross-links) using the respective $M_n$ and cross-linker densities using the following equations (8) & (9).[20]

$$\frac{1}{M_c} = \frac{2}{M_n} - \frac{\vartheta[\ln(1-\varphi)+\varphi+\varkappa\varphi^2]}{\vartheta_w(\varphi^{\frac{1}{3}}-\varphi/2)} \qquad (8)$$

$$\varphi = [1 + Q(\rho_p - \rho_w)]^{-1} \qquad (9)$$

Here, $M_n$ implies the number averaged MW of the polymer before cross-linking, $\upsilon$ and $\upsilon_w$ specific volumes of the polymer (1.22 g. cm$^{-3}$) and that of water (1 g. cm$^{-3}$), $\varphi$, the volume fraction of the polymer in swollen state, $\chi$, the interaction parameter is used as 0.45 for PAA-water.[20] Hence from the Q values and using the above relations it is possible to calculate $M_c$, the molecular weight between cross-links and the results have been plotted. (Fig. 2b). It is observed that the $M_c$ in case of TA cross-linked PAA increases till $x_{TA}$ to ca. 8.7 kDa and then gradually decreases to 7 kDa at higher STAG TA concentration (>$x_{TA}$). Likewise in case of U cross-linked PAA, $M_c$ increases to a maximum at $x_U$ and then decreases slowly to 4.7 kDa. On the other hand, $M_c$ remains almost constant hovering around 2 kDa for GD cross-linked PAA. (Fig. 2b)

Same trend is observed in case of the cross-linking density, $\rho_x$, which is calculated using the following relation. (Equation 10, Fig. 2c)[20]

$$\rho_x = (\upsilon M_c)^{-1} \qquad (10)$$

It clearly follows that the variation of the cross-linking density ($\rho_x$) with concentration in the cases of TA and U-STAGs reaches a maxima (at $x_{TA}$ and $x_U$) and then decreases. In case of GD cross-linked PAA the change of $\rho_x$ with GD concentration is almost constant. (Fig. 2c) However the general low Q-values and low 'solubility' for GD cross-linked PAA perhaps stems from the extensively tight network created by GD, insolubility of the network and above all very high thermal stability of the cross-linked polymer. The above observations of 'ʌ' type trend in the matter of variation of Q, $M_c$ and $\rho_x$ values with concentration in cases of TA and U-STAGs whereas low Q and almost constant $M_c$ and $\rho_x$ in case of GD cross-linked PAA draws attention to the other factor that might be operational in the swelling of the

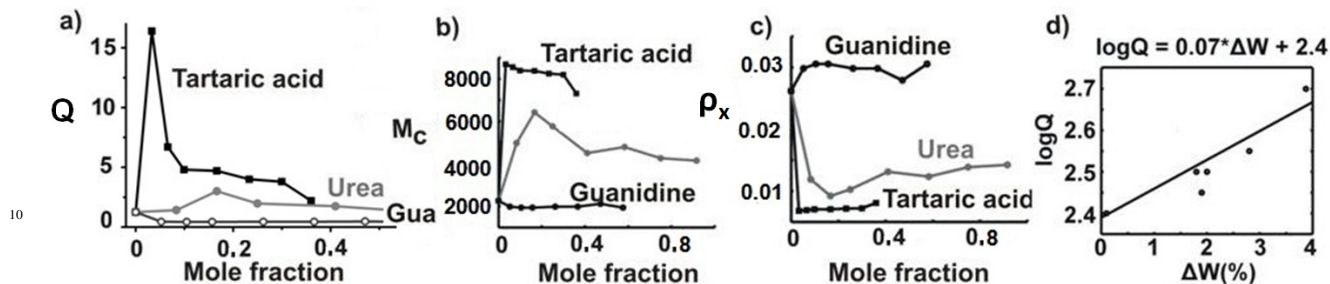

**Fig. 2** a) Swelling ratios (Q) for various mole-fractions of TA, U and GD used as STAG. (Note: the values of the mole fraction have been calculated from w/w% of STAGs used and was limited by their solubility in the PAA solution.) b) $M_c$ variation for the PAA cross-linked with STAGs according to equation (8) in the text. c) $\rho_x$ variation variation for the PAA cross-linked with STAGs according to equation (10) in the text. d) log Q plotted against the experimentally ascertained weight loss($\Delta W$ in %) from TGA values of various STAG cross-linked polymers to obtain the contribution of blowing in swelling of the STAG cross-linked polymers.

STAG networks: the effect of thermal aeration/blowing of the STAG networks.

**Correlation of thermal weight-loss of PAA-STAG-network polymers with swelling and its implication**

To find a clear effect, we studied the weight loss of the STAG network polymers in thermal decomposition studies and wanted to find a correlation of the weight loss with the Q values. To do so we heated the respective STAG-networks under air-flow emulating synthetic conditions in thermogravimmetric analysis (TGA) experiments and calculated the weight losses ($\Delta W$) at the processing temperature as given in the synthesis i.e., 150 °C. (See synthesis and experiments for details.) (Fig. 2d) From this study we find that although there is a clear correlation (equation 11 below) of the weight-loss ($\Delta W$) with swelling (or Q values) for TA and U-STAG networks, in case of GD there is no such weight loss. This observation in turn means GD is thermally inert and cannot aerate the network under the prevalent synthetic conditions. Being thermally stable GD can only act as a supramolecular cross-linker and the PAA-GD network becomes very highly cross-linked and hence insoluble or 'impermeable' to the incumbent water molecules. Such a factor leads to very low Q value and an almost constant $M_c$ and $\rho_x$ values for GD-PAA network. On the other hand in case of TA and U-STAGs network polymers the weight losses ($\Delta W$) of the networks at 150 °C in TGA experiments under air-flow and swelling (Q values) show a qualitative correlation. (Equation 11, Fig. 2d)

$$\log Q = 2.4 + 0.07\Delta W \\ = \log Q_c + 0.07 \Delta W \quad (11)$$

This correlation shows in addition to weight-loss ($\Delta W$) factor there is another constant contribution to swelling. We attribute that other contribution to supramolecular cross-linking and denote that contribution as $Q_c$. These results imply that for effective swelling there is interplay of two factors, viz., supramolecular cross-linking and thermal aeration. An optimum contribution of both factors lead to maximum swelling. This optimum interplay is achieved by variation of the concentrations of the STAGs in the cases of TA and U-STAGs. Moreover on a molecular level there has to be a constitutional level heterogeneity in disposition of TA and U-STAGs in the hydrogels to effect such variation of swelling as a function of their concentration. We speculate there are two types of U and TA: hydrogen-bonded and hence thermally stable TA and U, and non-hydrogen bonded and thus thermally labile TA and U in the network interstices. At lower concentrations or mole-fractions of U ($<x_U$) and TA ($<x_{TA}$) the non-hydrogen bonded STAGs in network interstices are depleted and the hydrogen bonded STAGs cross-link, till $x_U$ or $x_{TA}$ where the effect of thermal aeration and cross-linking reaches a synergetic maximum and yields highest swelling (or Q values). Beyond this concentration the effect of STAGs in cross-linking outweighs that of thermal aeration. This is because the available volume in the network interstices is limited and much lesser in number than the available hydrogen bonding sites of the entire network and hence the STAG-network tightens leading to a reduction in the overall Q value. This furthermore explains the 'Λ' type trend in the matter of variation of Q, $M_c$ and $\rho_x$ values with concentration in cases of TA and U-STAGs. (Fig. 2)

**Thermal aeration from microscopic investigations**

The effect of thermal aeration is also visible from SEM studies where we observe clear craters left open or porous structure formed in course of aeration in TA and U STAG networks whereas no such effect is observed in the pristine PAA network or the GD-PAA network vindicating our proposal of presence of a non-hydrogen bonding thermo-aero-able TA and U fractions in the corresponding STAG networks. Such pores seen in the SEM images of PAA-STAG-TA and PAA-STAG-U are due to the thermal decomposition of non-hydrogen bonded TA and U STAGs that leads to the evolution of gases like water vapour, CO, $CO_2$, and $NH_3$ respectively.[21,22] (Fig. 3) Decomposition of such non-hydrogen bonded TA, U, at 150 °C creating voids/craters, in turn increases the swelling ratios (Q) of the corresponding hydrogels. However for GD, the 'aero-ability' is minimum, which is also reflected in the non-porous SEM image of the PAA-STAG-GD. On the other hand, GD's capacity for hydrogen bond formation is rather high. Hence GD acts only as a supramolecular cross-linker among the STAGs and correspondingly PAA-STAG-GD shows lowest swelling ratio among all the hydrogels. Moreover it may be mentioned that at the processing temperature though inter- and intra-molecular condensation and chain transfer reactions in prime PAA network cannot be ruled out, such changes would affect all the STAG-networked PAA and pure PAA similarly and hence would not influence STAGs activity as reported here.

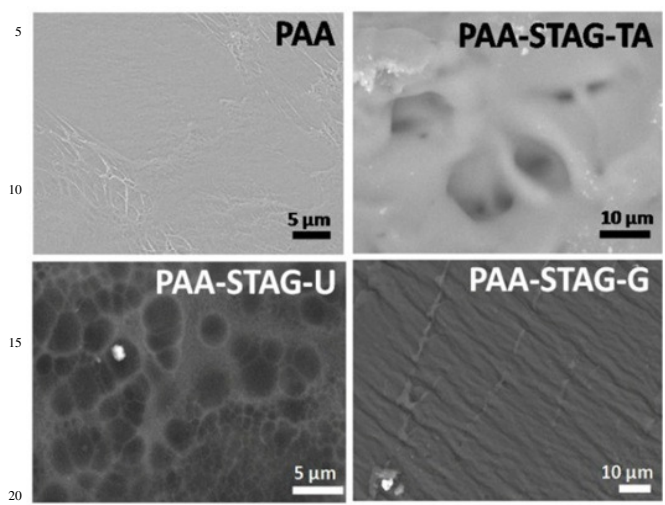

**Fig. 3** SEM images of PAA and those of PAA-STAGs clearly showing porous structures formed in the cases of TA and U.

### X-ray diffraction experimental investigations and amorphous nature of the gels

The structural aspects of these gels in dried state have been investigated with powder XRD. Dry powders of the gels were obtained by freeze drying the samples for 5 hrs. However, XRD studies do not show significant differences between PAA and various STAG cross-linked PAAs, like that of PAA-STAG-TA, PAA-STAG-U or PAA-STAG-GD. This might be due to the absence of any long range order and hence no information on the structure of the PAA-STAG hydrogels could be obtained. The powder X-Ray diffraction patterns shown in Fig. 4, do not reveal any significant difference. The XRD patterns of all the xerogels show typically features of amorphous polymers. Hence to obtain more structural information on these gels rather new Optical Diattenuance and Retardance studies were undertaken.

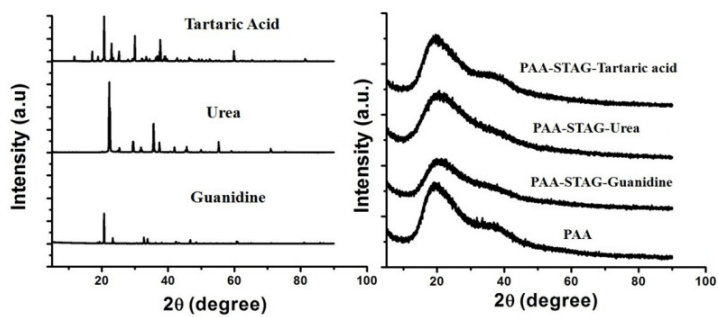

**Fig. 4** Powder XRD patterns of 3 STAGs (left); PAA and PAA-STAG xerogels (right).

### On structure of PAA and PAA-STAG-hydrogels from optical diattenuance and retardance studies: On origin of polarimetry effects in the PAA hydrogels reported in this study

Before we describe the results of the Mueller matrix analysis, the origin of the three polarimetry effects ($\delta$, $d$ and $\Delta$) in the samples, is worth mentioning. Linear retardance ($\delta$) arises due to difference in phase between orthogonal linear polarization states (between vertical and horizontal, or between 45° and -45°). Diattenuation ($d$) of any medium is due to differential attenuation of two orthogonal polarization states. Note that at macroscopic level, $\delta$ and $d$ have a common origin. Both effects arise from differences in refractive indices (real and imaginary parts respectively) for different polarization states. They are often described in terms of ordinary and extraordinary indices and axes. Thus, anisotropically organized structures stemming from their molecular anisotropy/organization often manifest as linear retardance or diattenuation effects. Hydrogels, like the ones reported here, in general may exhibit linear retardance due to organization of individual anisotropic domains. Such retardance originates from anisotropic molecular polarizability. The net value of $\delta$ is determined by a number of factors. They are: the molecular structure, the value of (anisotropic) polarizability, ease of queuing up monomers and the of course synthetic conditions, are few among many such factors.

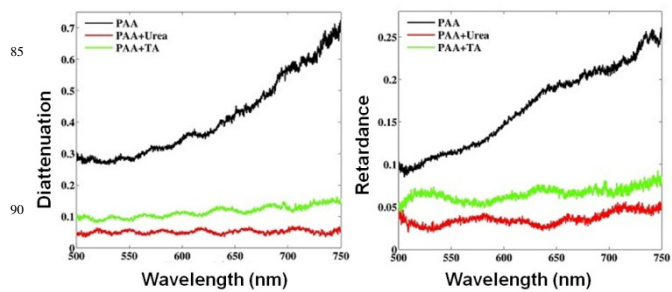

**Fig. 5** Wavelength variation of Diattenuance (left) and Retardance (right) for pure PAA, PAA-STAG-TA and PAA-STAG-U.

The effect of diattenuation ($d$) is analogous to retardance. It is also determined by organization and molecular structure. The parameters affecting the net diattenuation are as following. (i) Anisotropic absorption from oriented/organized structure. (ii) Scattering, reflection/transmission of light through layered structures. The remaining polarimetry effect, $\Delta$, is primarily caused by following factors. (i) Multiple scattering effects within the sample (due to turbidity) and (ii) randomly oriented spatial domains of birefringent structures.

### Electronic absorption spectroscopy (EAS) on hydrogels & its influence in diattenuance studies

The hydrogels investigated in our study did not show appreciable absorption in the wavelength range 500 – 750 nm (Fig. 6) and thus the latter effect i.e., scattering, reflection and transmission of light through organized layered structure is expected to be the dominant contributor to the observed diattenuation of the samples. Their electronic absorption spectra are given below. The rather sharp increase in intensity below 300 nm is indicative of scattering by the hydrogel as mentioned above.

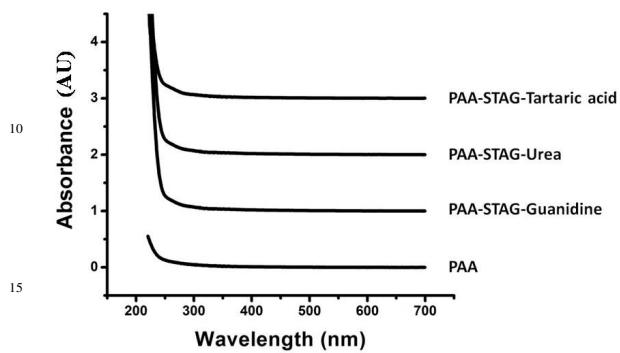

**Fig. 6** Electronic absorption spectroscopic studies of the hydrogels.

In Fig. 5, we show the wavelength variation ($\lambda$ = 500–750 nm) of the Mueller matrix decomposition-derived diattenuation ($d$) and retardance ($\delta$) parameters, respectively, for pure PAA hydrogel and two other PAAs with STAGs Tartaric acid (TA) and Urea. In general, the sample of pure PAA exhibits much stronger diattenuation and linear retardance effects compared to the other two mixed samples (with TA and urea added to PAA). The results indicate the following two features. (i) The reduction of linear retardance $\delta$ with added urea and TA, in general are indicative of randomization of the organization of the oriented anisotropic micro-domains inside those PAA-STAG-TA and U hydrogels. (ii) As mentioned earlier, diattenuation in these hydrogels primarily originate from their layered structures. Thus reduction of the magnitude of diattenuation in the U and TA added hydrogels indicates that the layers in these two hydrogels are organized in a more relaxed fashion and the layers are thus expected to be more separated in both these gels as compared to the pure PAA or the one with guanidinium hydrogel.

**On the dynamics of microstructure of PAA-STAGs from variable temperature $^1$H NMR spectroscopy**

VT $^1$H NMR studies were undertaken to observe the change in dynamics in the microstructure of the PAA-STAG hydrogels. In all cases the acquisition is difficult due to gelation. However following features are discerned and are explained here. In all the 4 cases of the gelled samples, viz., PAA, PAA-STAG-TA, PAA-STAG-U, PAA-STAG-GD, there is a very strong peak around 4.8 ppm (Fig. 7). This is attributed to the free water in the gel.[23] The peaks around 2 ppm is attributed to the methine and methylene protons of PAA.[23] The peak around 2 ppm at 20 $^o$C is rather broad. This implies that the sample is totally hydrated. The heavy hydration leads to high mobility of the PAA chains in all the cases and leads to peak broadening. However, on heating the peak sharpens and narrows down. This implies dehydration. Such dehydration results in a decrease of the mobility of the PAA chains. This results in an increase in the dipolar interaction between PAA main chain protons and hence leads to sharpening of the peaks. This feature is uniform in all the cases.

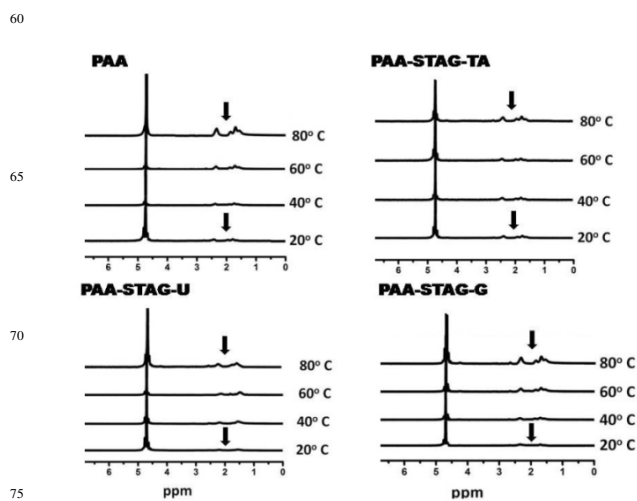

**Fig. 7** Variable temperature $^1$H NMR spectra of the hydrogels. The black arrow shows the sharpening of the peak corresponding to methylene protons with increasing temperature.

**On supramolecular cross-linking by STAGs from FTIR spectroscopy**

The probes were also investigated by FTIR spectroscopy in dried state. Distinct differences in spectral signatures were observed between the starting PAA and PAA-STAG networks implying effective supramolecular cross-linking in the latter. In all cases, the PAA-STAG networks showed a blue shift in the spectral signature of the carboxylate, amide or imine bonds, implying a stiffening of the corresponding bonds due to hydrogen bonding in the PAA-STAG networks (Fig. 8 and Fig. 9). More precisely, in case of TA cross-linked PAA it is observed that a single intense peak is observed at 1620 cm$^{-1}$.[22]

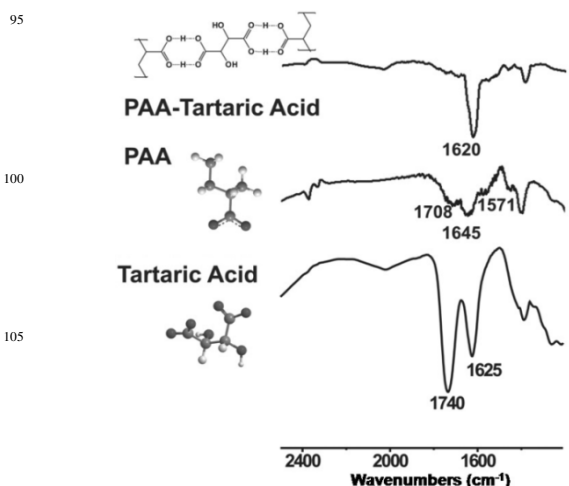

**Fig. 8** Comparative FTIR spectra of PAA and PAA-STAG-TA. This is due to intermolecular hydrogen bonding of the C=O groups in the TA cross-linked polymer.[24] It is further interesting

to note that the stretching frequencies of the intramolecularly hydrogen bonded C=O at 1645 cm$^{-1}$ in case of both pristine PAA and that of 1645 cm$^{-1}$ in TA are blue shifted to 1620 cm$^{-1}$ in the PAA networked with TA STAGs. The above shifts thereby show the presence of a more rigid hydrogen bonded/supramolecularly connected network.[24] Moreover the asymmetric stretching vibration of COO$^-$ in the pure PAA at 1570 cm$^{-1}$ disappears due to cross-linking in all PAA-STAG networks. This is because such stretching modes are not allowed in a network. Likewise the reduction in intensity of the peak at 1708 cm$^{-1}$ corresponding to not hydrogen bonded CO and COO$^-$ stretch in PAA is reduced in case of PAA-STAG-TA. This implies that in case of the PAA-STAG-TA, the carboxylate group is not free but is supramolecularly linked to another species, i.e., TA. Similar patterns of shifts in stretching frequencies follow from U and GD cross-linked PAA. For instance in case of (free) U we observe a H-free NH stretching vibration at 3450 cm$^{-1}$ and an amide I band at 1640 cm$^{-1}$ along with a $\delta_{NH}$ vibration at 1450 cm$^{-1}$. In the U STAG PAA, it is observed that a blue-shifted NH stretching band appears at 3250 cm$^{-1}$ in addition to the other characteristic spectral signatures of U and PAA. Such a blue shift stems from stiffening of the corresponding bond due to an effective cross-linking. Similarly a blue shifted amide I band at 1620 cm$^{-1}$ is also observed in the PAA-cross-linked U in addition to a blue-shifted $\delta_{NH}$ vibration at 1384 cm$^{-1}$. (Fig. 9) These shifts imply that there is an effective hydrogen bonding/supramolecular cross-linking by U in the U STAG PAA-network. Similar spectroscopic evidences are obtained for GD as well.

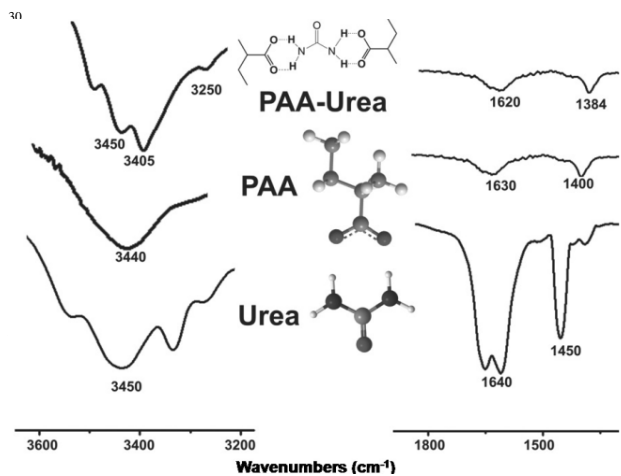

**Fig. 9** Comparative FTIR spectra of PAA and PAA-STAG-U.

## Conclusions

To summarize: we have shown here that it is possible to synthesize supramolecularly cross-linked polymeric hydrogels (PAA-based) using STAGs like TA and U. The GD on the other hand being an efficient supramolecular cross-linker and thermally rather stable forms a highly cross-linked insoluble polymeric network with very low swelling in water. GD in this study acts as a contrasting reagent to show the efficiency of STAGs like TA and U. The STAGs reported here can reduce significantly the cost of hydrogel-production and are in a sense greener alternative to the existing routes as they are bi-functional and hence alleviate the need of two separate components (cross-linking and blowing).

The authors thank Dr. Nirmalya Ghosh and Jalpa Soni of the Department of Physical Sciences, IISER-Kolkata for their help with optical diattenuance and retardance studies. The authors thank Natural Science Foundation, China, CIT, and IISER-K for financial support. AS, DDY and FFX contributed equally to this work. This work is dedicated to Prof. Bruno Chaudret.

## Notes and references

[a] School of Chemistry and Materials Engineering, Changshu Institute of Technology, Changshu, Jiangsu, P. R. China. Fax: +8651252251821
[b] Eco-Friendly Applied Materials Laboratory, DCS, New Campus, IISER-Kolkata, India.
Fax: +913325873020, *E-mail: roy.soumyajit@googlemail.com, s.roy@iiserkol.ac.in (SR), yuanrx@cslg.cn (RY).

**Table of Contents Entry**

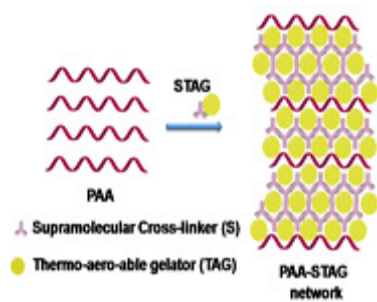

Supramolecular Thermo Aero-able Gelators (STAGs): tartaric acid, urea, and guanidine with amide and imine moieties as supramolecular synthons are reported that simultaneously cross-link and aerate ('aero-able') polyacrylate networks for synthesis of hydrogels.